\documentstyle[epsfig]{article}
\def\jpsi{\mbox{J/$\Psi~$}}
\def\kts{\mbox{$k_{T}^{2}$}}
\begin{document}
\begin{titlepage}
\hspace*{\fill}{IMSc-96/11/29}
\vspace*{\fill}
\begin{center}
{\Large \bf Validity of double scaling analysis in semi-inclusive
processes - $J/\psi$ production at HERA}
\\[1cm]
Tapobrata Sarkar and Rahul Basu
\footnote{email: sarkar,rahul@imsc.ernet.in}\\
{\em The Institute of Mathematical Sciences, Chennai (Madras) 600 113, India}
\end{center}
\vspace{2cm}
\begin{abstract}
In this paper we check the validity of the ideas of double 
scaling as given by Ball and Forte in a
semi inclusive process like $J/\psi$ production at HERA, in different 
kinematical regions, for low values of the Bjorken variable $x$. In particular, 
we study $J/\psi$
production in the inelastic and diffractive (elastic) regimes using the double
scaling form of the gluon distribution functions. We compare these predictions 
with data (wherever available) and with other standard parameterisations. 
We find that double scaling holds in the inelastic regime over a 
larger kinematic region than
that given  by the analysis of the proton structure function  $F_2^p$. 
However, in the diffractive region, double scaling seems to suggest an 
admixture of hard pomeron boundary conditions for the gluon distribution, while
predicting a steeper rise in the cross section than suggested by present data. 
\end{abstract}
\vspace*{\fill}
\end{titlepage}
\section{Introduction}
\par
The systematic study of the dynamics of perturbative QCD at low values
of the Bjorken variable $x$ has thrown up many new ways of analysing the
structure of the proton through a measurement of the structure function
$F_2^p$. In particular, one of the methods that provides a clean study
of the scaling properties of structure functions as $Q^2$ (the
virtuality of the photon) increases and $x$ decreases is the
so-called double scaling analysis of Ball and Forte \cite{bf}. In their
work, in the kinematic region of very large values of the
photon-proton centre of mass energy squared, $S$, the Altarelli-Parisi
(DGLAP) evolution equation were solved and it was shown that the gluon
distribution (which is the dominant partonic distribution in the region
of interest) shows double scaling in terms of the variables 
\begin{equation}
\sigma\equiv\sqrt{\ln\frac{x_0}{x}\ln\frac{t}{t_0}}\ \ ;
\rho\equiv \sqrt{\frac{\ln x_0/x}{\ln t/t_0}}
\end{equation}
where the starting scale for $Q_{0}^{2}$ in
$t_0\equiv \ln \frac{Q_0^2}{\Lambda^2}$ can be
just a little more than $Q_0^2=1 GeV^2$,  $\Lambda= \Lambda_{QCD}$,
$x_{0}=0.1$, typically.
This asymptotic analysis is a straightforward consequence of
perturbative QCD evolution of structure functions at low $x$, and
provides excellent agreement with the data for $Q^2 \geq 5 GeV^2$.
However, taking into account the fact that other analyses, the most
common being the BFKL approach \cite{bfkl}, also provide a mechanism for
predicting the behaviour of $F_{2}$ at low $x$, a case has been made for
studying the validity of some of these different approaches in processes
that are not as inclusive as $F_{2}$. One such process where the
validity of the double scaling hypothesis has not been tested is \jpsi
production. In this paper, we propose to remedy this by studying
\jpsi production at high centre of mass energies of the kind
available at HERA, using the double scaling form for the gluon
distribution function.

The issue of \jpsi production at large centre of mass energies 
available at HERA (${\sqrt S} \sim 200$GeV) has become an important aspect of
QCD because, generally, production of heavy quarkonium states in 
high-energy collisions is an important tool for the study of its various
perturbative and non-perturbative aspects. The production of heavy quarks in a
hard scattering process can be calculated perturbatively. However the
subsequent creation of a physical bound state (\jpsi or other quarkonia)
involves non-perturbative QCD. This latter process is treated in a factorised
approach where the short and long distance physics are separated out. In
particular, the evolution of a quark-antiquark pair into a physical
quarkonium state is described by the color singlet or color octet models to
which we will return in more detail later. 

In this paper, we shall address the issue of \jpsi production at low values of 
the Bjorken variable $x$.
At low values of $x$, the gluon distribution plays a dominant 
role in the determination of the cross section as mentioned earlier. 
In particular, we concentrate
on the specific kinematic region of very large values of the
photon-proton center of mass energies, $\sqrt S$, in e-p collisions.
We calculate the \jpsi production cross sections at low $x$, 
using the asymptotic form for the gluon distribution function, given by the
(leading order) double scaling analysis in 
different kinematical regimes, namely for inelastic 
and diffractive scattering. 
Till now, most calculations for \jpsi production use the standard
parametrisations for the gluon distributions, based on a fit to the known data. The
advantage of using the double scaling form is that it is, in a certain sense, 
derived from QCD and has no fitting parameters other than the starting values 
of $Q^2$ and
$x$ from where the distributions are evolved. We have, in fact, checked the
dependence of our predictions on these starting values in order to be able to
quantify the degree of uncertainty that arises from these parameters. 
for diffractive \jpsi photoproduction.
\par
The paper is organized as follows. In section $2$, we briefly
summarize the phenomenon of double asymptotic scaling of the gluon
distribution function. In section $3$, we present our results for \jpsi
production in the inelastic regime. We then, in the next section move on
to the elastic regime, where BFKL dynamics is supposed to be dominant
and make our predictions regarding the cross sections. 
Although it may seem inconsistent to use the gluon distribution
function obtained by solving the Altarelli-Parisi evolution equation
in the double scaling limit in the diffractive
regime where BFKL dynamics takes over, it may be pointed out
that these distribution functions are merely used as parametrizations
of the data. 
\par
In the last section 
we discuss our results and various open problems.
\par
All the experimental data included in this paper have been obtained from
\cite{aid} except where otherwise stated.
\section {Gluon Distribution at low x}
\par
There have been two distinct approaches to the study of the HERA
data, the 'standard' Altarelli Parisi (DGLAP) \cite{ap} 
evolution equation approach and the 
attempt to reconcile the data with BFKL dynamics \cite{bfkl}. 
In order to study the effects of summing $\alpha_s\ln {1\over
x}$ unaccompanied by $\ln Q^2$, the preferred approach has been to
study the BFKL equation which generates a singular $x^{-\lambda}$
behaviour for the unintegrated gluon distribution $f(x,k_T^2)$, with
$\lambda={\alpha_s} 4\ln 2$ (for fixed, not running $\alpha_s$).
For running $\alpha_s$, $\lambda\simeq 0.5.$ \cite{bfkl}. From this an
asymptotic form for $F_i(i=2,L)$ can be inferred.
We shall come back to the issue of BFKL dynamics in more detail when we
deal with diffractive \jpsi photoproduction, later in the paper.
In the other approach one attempts to describe the data through
the DGLAP evolution equation to the
next-to-leading order approximation. Here again, the data imply a
steep gluon distribution with the gluon density rising sharply as $x$
decreases, even for comparatively low values of $Q^2$. Ball and Forte
\cite{bf} have used this approach to exhibit
the scaling properties of the gluon distribution function and hence the
proton structure function $F_{2}$
at low $x$, generated by QCD effects and have
shown that the HERA measurement of $F_2(x,Q^2)$ is well explained by
this approach.
In \cite{bf} it was shown that the solution of the traditional
DGLAP equations in the low $x$ regime exhibits double
scaling in terms of the variables defined in (1).
Double asymptotic scaling
results from the use of the operator product expansion and the
renormalisation group at leading (and next-to-leading) order, and
predicts the rise of $F_2$ on the basis of purely perturbative QCD
evolution.
The asymptotic behaviour of the gluon density in this approach is given
(to leading order) by
\begin{equation}
xg(\sigma,\rho) \sim {N \over \sqrt {4 \pi \gamma \sigma}} exp \left[ 2 \gamma
\sigma~-~\delta ({\sigma \over \rho}) \right] \left( 1+O({1 \over
\sigma}\right)
\end{equation}
which leads to the asymptotic 
behaviour of $F_{2}$:
\begin{eqnarray}
F_2^p(\sigma,\rho)&\sim& N\frac{\gamma}{\rho}
\frac{1}{\sqrt{4 \pi\gamma\rho}}exp[2\gamma\rho-\delta(\frac{\sigma}{\rho})]
\nonumber \\
&&\times [1+O({1\over\sigma})]
\end{eqnarray}
where
$$
\gamma\equiv 2\sqrt{\frac{n_c}{\beta_0}} \ \; \beta_0=11-{2\over
3}n_f
$$
and
$$
\delta\equiv (1+\frac{2n_f}{11n_c^3})/(1-\frac{2n_f}{11n_c})
$$
$n_{c}$ and $n_{f}$ being the number of colors and the number of flavors
respectively.
$N$ is a constant that depends on the input gluon distribution. In our
analysis, we have taken $N=3.24$, corresponding to soft boundary 
conditions on the gluon \cite{bf} as favoured by HERA data \cite{aid1}, in 
the region of $x$ that we are interested in, for inelastic \jpsi
production. For \jpsi photoproduction in the diffractive scattering
regime, we shall see that harder boundary conditions are preferred and
in this kinematical region, we have taken an intermediate pomeron,
corresponding to $N=1.45$.
\par
In our calculation, for the \jpsi production cross sections in this
double scaling limit, at small values of $x$,
we have used the values $Q_{0}^{2}=1.12 GeV^{2},
\Lambda=0.248GeV$ and the starting value of $x_{0}=0.1$.
For the values of the variables $x$ and $Q^{2}$ for which
$\sigma^{2}>1$ and $\rho^{2}>2$, double scaling of the proton structure
function $F_{2}$ is confirmed to a fair degree. 
However, in the semi-inclusive \jpsi production process, we shall not
restrict ourselves to these kinematical regimes, but rather try to 
check the region
of validity for double scaling where data is available.
\section{The Inelastic Region}
In the color singlet
model approach, the dominant contribution to the quarkonium production
comes from quark-antiquark or gluon-gluon fusion. The fusion process
produces a heavy quark pair which forms a physical color singlet state
with an appropriate color singlet projection. 
\begin{figure}[htb]
\begin{center}
\mbox{\epsfig{file=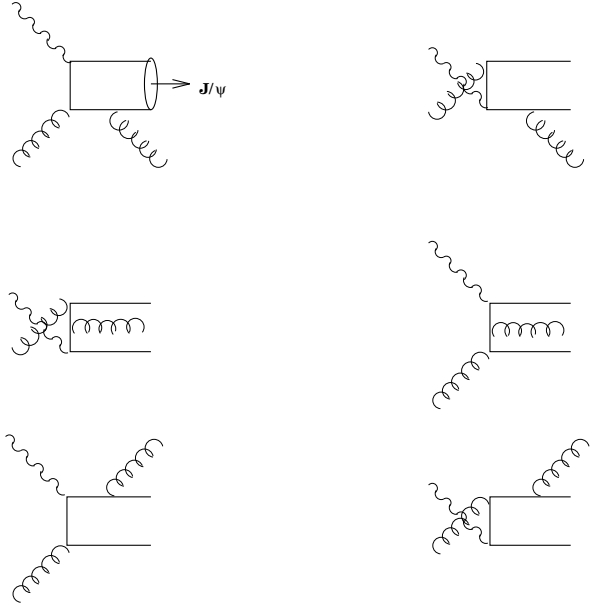,width=8truecm,angle=-90}}
\caption{ Color Singlet contribution to inelastic \jpsi production.}
\label{Fig 1.}
\end{center}
\end{figure}
This fusion contribution to J/$\Psi$ production in the color singlet
model has been computed in \cite{B/J}. It takes place via the subprocess
$\gamma g \rightarrow J/\Psi g$ where the gluon in the final state
of the partonic scattering process carries off the color charge. The final
$c \bar c$ state is a color singlet and the cross section for the process
$\gamma N \rightarrow J/\Psi X$ can be calculated from the above subprocess 
and is given by
\begin{equation}
{d^{2}\sigma \over dtdx}
~=~{8 \pi \alpha_{s}^{2} \Gamma^{J}_{ee}m_{J}^{3} \over 3 \alpha}
g(x)f(s,t)
\end{equation}
where,$$ f(s,t)~=~{1 \over s^{2}}\left[ {s^{2}(s~-~m_{J}^{2})^{2}
~+~t^{2}(t~-~m_{J}^{2})^{2}~+~u^{2}(u~-~m_{J}^{2})^{2} \over 
(s~-~m_{J}^{2})^{2}~(t~-~m_{J}^{2})^{2}(u~-~m_{J}^{2})^{2}} \right]
$$
with $s~+~t~+~u~=~m_{J}^{2}$, and $\Gamma^{J}_{ee}$ the electronic width
of the $J/\Psi$ which is related to its orbital wave function at the
origin ($\Gamma^{J}_{ee}=5.26KeV$ in our calculations).  
Here, $s,t,u$ are the usual Mandelstam variables for the
partonic subprocess and $s$ is related to the total centre of mass
energy $S$ by $s~=~xS$ ,$x$ being the fraction of the incident nucleon
momentum carried by the gluon.
It is more convenient to re-express this differential cross section in
terms of the variables $z$ the elasticity parameter, and $P_{T}^{2}$,
the transverse
\jpsi momentum squared.
The elasticity parameter $z$ is defined by 
$$ z~=~{P_{J/ \psi} .P_{N} \over Q.P_{N}}$$
where $Q~=~P_{\gamma}~=~$ momentum of the incident virtual photon.
$z$ denotes the fraction of the energy transferred to the \jpsi in the
process and in particular in fixed target experiments, reduces to 
$z~=~E_{J/ \psi}/E_{\gamma}$
\par
In terms of $z$ and $P_{T}$, we may write
the Bjorken variable as
\begin{equation}
x~=~{1 \over S} \left[ {m_{J}^{2} \over z}~+~{P_{T}^{2} \over z(1~-~z)}
\right]
\end{equation}
It is more transparent to reexpress this differential cross section in
terms of the variables $z$ and $P_{T}^{2}$. In terms of these variables, the
differential cross section is given by,
\begin{equation}
{d^{2} \sigma \over d P_{T}^{2} dz}~=~{xg(x)z(1~-~z)m_{J}^{4}A \over 
[m_{J}^{2}(1-z)+ P_{T}^{2}]^{2}} {\bar f}(z, P_{T}^{2})
\end{equation}
where, $A~=~{8 \pi \alpha_{s}^{2} \Gamma_{ee}^{J} \over 3 \alpha m_{J}}$ 
and the function ${\bar f(z, P_{T}^{2})}$ is given by 
\begin{equation}
\bar f(z, P_{T}^{2})~=~{1 \over (m_{J}^{2}+ P_{T}^{2})^{2}}~+~{(1-z^{4}) \over
(P_{T}^{2}+m_{J}^{2}(1-z)^{2})^{2}}~+~{z^{4}P_{T}^{4} \over (m_{J}^{2}+
P_{T}^{2})^{2}(P_{T}^{2}+m_{J}^{2}(1-z)^{2})^{2}}
\end{equation}
This equation can be used to evaluate the differential cross sections ${d
\sigma \over d P_{T}^{2}}$, ${d \sigma \over dz}$ 
and also the $P_{T}^{2}$ integrated total cross
section. It may be remarked here that the above expressions show that
the cross sections are finite for non zero $P_{T}^{2}$
for all values of $z$,
and for $P_{T}^{2}=0$,
it diverges in the limit when $z=1$. However,
this value
of $z$ is excluded from the domain of our interest, as 
there is a strict upper bound on $z$ for this inelastic process, {\em
viz.} the
restriction that the momentum transferred in the partonic subprocess be
above the QCD scale so that perturbative QCD is valid; this restriction,
$|t| \geq Q_{0}^{2}$ gives 
\begin{equation}
(1~-~z)~>~{Q_{0}^{2}- P_{T}^{2} \over Q_{0}^{2}+m_{J}^{2}}
\end{equation}
and hence, corresponding to $Q_{0}^{2} \simeq 1.12 GeV^{2}$, we have $z
\leq 0.90$. 
\par
%
Before we present the results of our calculation, a word about the
applicable region of phase space for the double scaling form
of the gluon distribution function is in order. 
We have already remarked that
the predictions of the proton structure function $F_{2}^{p}$ at HERA shows 
that asymptotic behaviour sets in, in terms of the variables $\sigma$ and
$\rho$ defined in the beginning, for $\sigma^{2}~>~1$ and $\rho^{2}~>~
2$. As can be seen from the expression for $x$, imposing the cuts 
$ z \leq 0.9$, we cannot really go to sufficiently small values of
$x$ for large values of $P_{T}^{2}$, which is essential for the validity
of the asymptotic form of the gluon distribution.  
To be well within the 
double scaling region in $x$ and $Q^{2}$, for such large values of 
$P_{T}^{2}$, one thus has to impose much stronger cuts on $z$.
However, we shall not use such cuts in our analysis as we try to go
beyond the limits for $\sigma$ and $\rho$ as obtained from the analysis
of $F_2^p$.
\begin{figure}[htb]
\begin{center}
\mbox{\epsfig{file=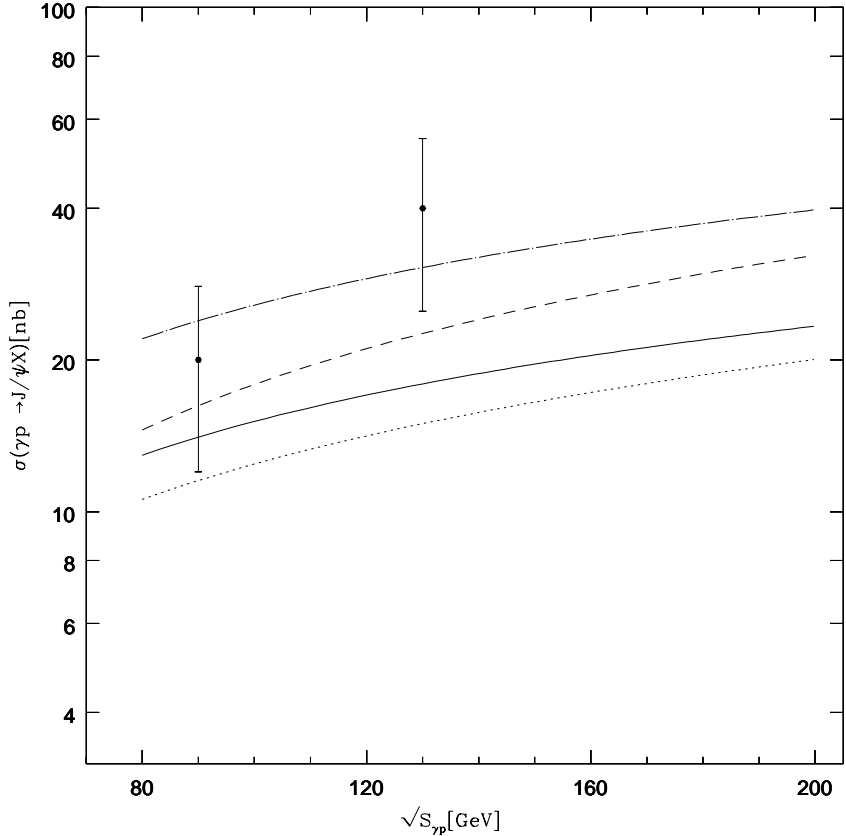,width=12truecm}}
\caption{ Total cross section for inelastic \jpsi production.
The solid line corresponds to the Ball and Forte gluon distribution, 
the dashed line to the GRV and the dotted line to the CTEQ
parametrisations. The dashed-dotted line shows a fit to the data with
the Ball and Forte distribution assuming a K-factor of 1.7.}
\label{Fig 2.}
\end{center}
\end{figure}
\par
We have analysed the behaviour of the total cross section and the
$P_{T}$ and z differential cross sections for \jpsi production. The results 
of the analysis are shown in Figs.(2),(3),(4).  
In our calculation, we have taken $m_{c}=1.4$ and 
the starting scale $x_{0}$ is taken to be $0.1$. 
For the calculation of the total cross section, we have imposed the
cuts $0.1 \leq z \leq 0.8$ and $P_{T}^{2} \geq 1GeV^{2}$,
whereas for the $P_{T}$ differential cross section, the cut on the
elasticity paramter has been imposed as $0.1 \leq z \leq 0.9$ in order to match
with experimental data from \cite{aid}. Cuts on $P_{T}$ are 
generally imposed as in the region of small $P_{T}$, the non leading order
calculations are not fully under control. Also, the lower bound on $z$ is 
required in order to suppress contribution to \jpsi production arising out of 
B decays.
We have checked that on inclusion of a K factor of $1.7$ 
in the total cross section data (as in \cite{kram}), the
results are in fair agreement with extrapolated data points \cite{aid}.
From the results it is also clear the one can extend the double scaling
validity region to kinematic ranges that go beyond that obtained from
the analysis of the proton structure function $F_2$. Indeed, one gets
fairly good agreement with the data for values of the variable 
$\sigma^2 \geq 0.4$, $\rho^2$ being typically greater than $2$. 
\par
We should mention here that in the expression for the total cross
section and $d \sigma / dz$, one has to in principle integrate over all
values of $P_{T}^{2} \geq 1 GeV^{2}$. 
From the expression for $x$, it is clear
that to remain at low values of $x$, 
one cannot really go to large
values of $P_{T}$ ($P_{T} \geq 6 GeV$, say), without severely
constraining the inelasticity parameter $z$, as we had pointed out. 
However, as can be seen from the
expression for the cross section, in the approximation that $P_{T}^{2}$ is
large compared to $m_{J}^{2}$, the denominator falls off as $P_{T}^{8}$,
and hence the contributions from the large $P_{T}$ region is negligible
for such large values of $P_{T}$. 
Hence, in our numerical calculations, we have restricted ourselves
to $P_{T} \leq 6GeV$.

The $z$ and $P_{T}^{2}$ distributions are shown in Figs(3) and (4).
From the $P_{T}^{2}$ distribution, it is seen that the double scaled form
of the gluon distribution function underestimates the data to some
extent, however, the K factor (enhancement factor for the cross sections
on inclusion of the NLO corrections)
might take this into account at
least in part. In the calculation of the differential cross sections
also, we have, as mentioned earlier
restricted ourselves to the region $P_{T}^{2}
\geq 1 GeV^{2}$ as in the limit of extremely small transverse momentum,
there are large negative contributions coming from NLO corrections and
the cross section falls sharply as $P_{T} \rightarrow 0$.
\par
\begin{figure}[htb]
\begin{center}
\mbox{\epsfig{file=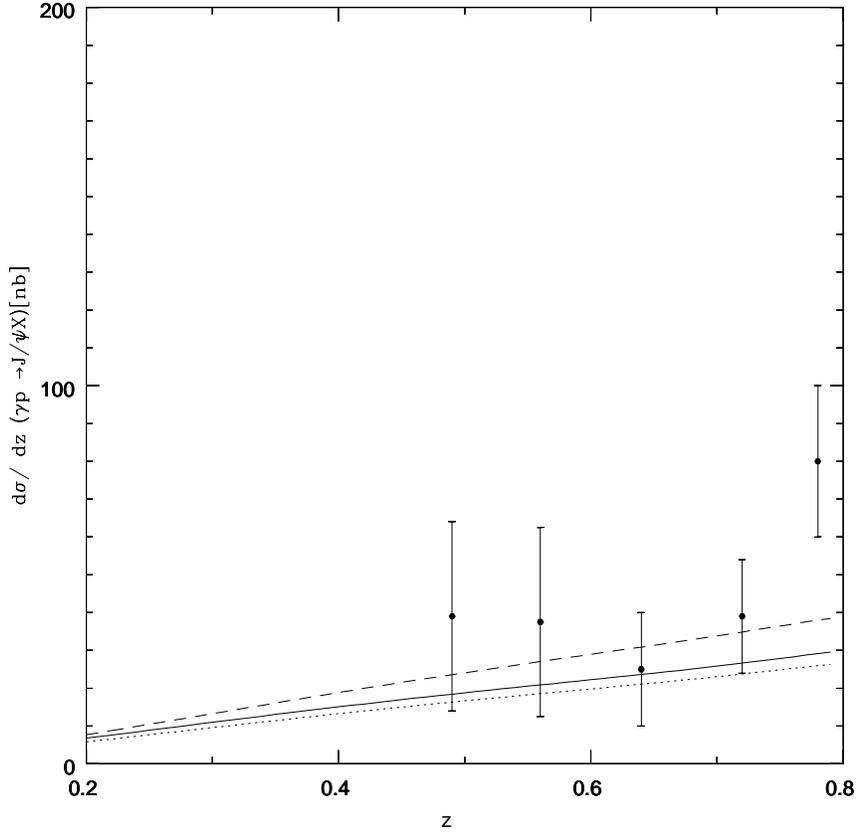,width=12truecm}}
\caption{ z Differential cross section for inelastic \jpsi production
at $\protect\sqrt {S_{\gamma p}}=100GeV $
The solid line corresponds to the asymptotic gluon distribution, the
dashed line to the GRV fit and the dotted line to the CTEQ
parametrisation.}
\label{Fig 3.}
\end{center}
\end{figure}
\begin{figure}[htb]
\begin{center}
\mbox{\epsfig{file=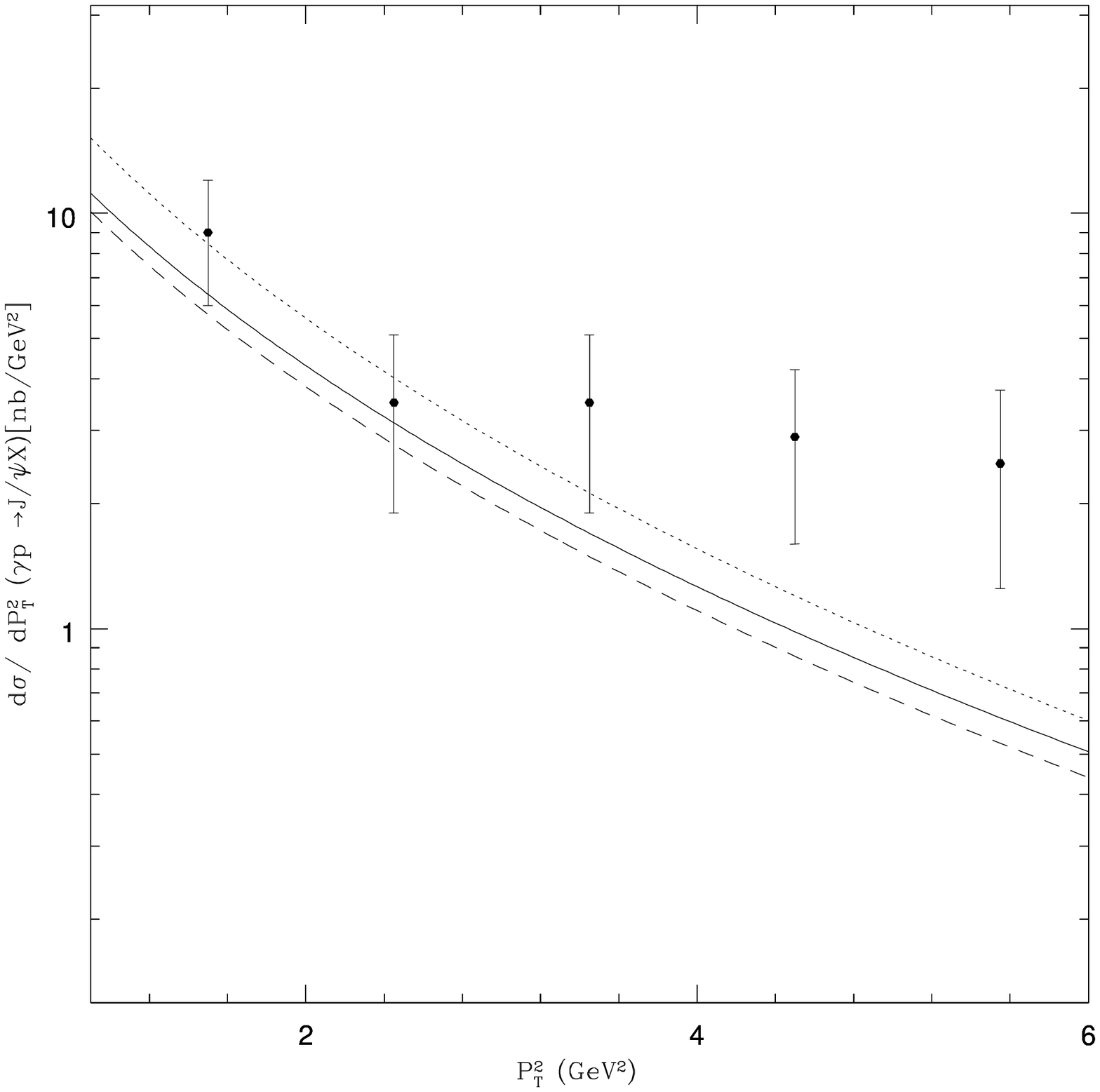,width=12truecm}}
\caption{ $P_{T}^{2}$ Differential cross section for inelastic \jpsi 
production at ${\protect\sqrt S_{\gamma p}}=100GeV $.
The solid line corresponds to the Ball Forte distribution, the dotted
line to the GRV and the dashed line to the CTEQ parametrisations.}
\label{Fig 4.}
\end{center}
\end{figure}
In our calculations, we have not considered "color octet" 
contributions to the J/$\Psi$ cross sections.
In the color singlet model, that we are considering, 
one conventionally assumes that the color
singlet $c \bar c$ state is produced in the $~^{3}S_{1}$ state initially
and then evolves into the physical \jpsi state. This 
model however grossly underestimates the \jpsi production data at the
Tevatron \cite{mangano}. To reconcile with the data, a new mechanism for
\jpsi production was suggested in \cite{bratflem}, the color octet
mechanism.
Whereas in essence the "color
singlet" model ignores the relative velocity $v$ between the quarks in
the bound state, this quantity is not often negligible and may give rise
to considerable amount of corrections to the results. In \cite{bbl} 
an expansion of the quarkonium wave function in powers of $v$ and 
corrections involving the octet components of the wave
function have been carried out. However, as has been pointed out in
\cite{cacc} color octet contributions may become important only in the
diffractive limit, i.e $z$ close to $1$ , for small $P_{T}$. It was
further demonstrated in \cite{cacc} that at HERA experiments, 
in the inelastic region, i.e for $z \leq 0.9, P_{T}^2 \geq 1$, the
data can be well accounted for by inclusion of the next to leading
order corrections to the cross sections, and hence 
the color octet contributions in these energy ranges do not seem to
play a very significant role. Comparison of the color singlet and color
octet contributions to ${d \sigma \over dz}$ with experimental data has
been carried out in \cite{cacc} and this analysis shows that the
\jpsi spectrum is well accounted for by the color singlet model alone
and the octet model shows a marked increase in cross section at large
values of $z$, which is clearly in conflict with experimental data.
Hence we are justified in dealing only
with the color singlet model in our calculations. 
\par
The choice of scale as $Q~=~P_{T}/2$ would have in
part accounted for the K factor mentioned earlier; 
however, with this choice of scale, small values of
$P_{T}$ in the double scaling regime cannot be reached, as 
from the form of the gluon distribution function (2) defined in terms of
the variables $\sigma$ and $\rho$ in (1), it is clear that it diverges
for $Q=Q_0$.
We have thus chosen our scale as
the factorization scale, namely $Q^{2}=2m_{c}^{2}$.
\par
Also, at HERA energies, contributions to the \jpsi production cross
section might arise from B decays, however these start contributing
significantly at extremely low values of $z$, namely $z \leq 0.1$
\cite{ng}
We have ignored such small values of $z$ in our analysis, hence this
effect will have negligible effect on our results.
\par
Before ending this section, 
we would like to point out that although we have considered here only the 
fusion process contribution to \jpsi production, at large values of
$P_{T}$, 
the fragmentation process, i.e the fragmentation of gluons and
charm quarks, also become an important source of \jpsi production.
Even though these contributions appear at higher orders of the strong
coupling as compared to the fusion contributions, they become important
at large values of the \jpsi transverse momenta due to an enhancement
factor of $P_{T}^{2}/m_c^{2}$, $m_c$ being the charm quark mass.
Consequently, fragmentation process contributions may dominate over
fusion at comparatively large values of $P_{T}$.
Fragmentation functions are computed perturbatively, at
the initial scale of the order of the charm quark mass and the
Altarelli-Parisi equations are then used to resum large logarithms in
$P_{T}/m_{c}$. Gluon and charm fragmentation functions have been
calculated \cite{B/Y},\cite{B/Y1} and it has been shown 
\cite{RMG} that at HERA experiments, the
charm quark fragmentation process is expected to dominate over the
quark gluon fusion process at sufficiently values of $P_{T} \geq 7GeV$.
Calculation of the fragmentation functions in the double scaling regime
remains an open issue and we hope to address this elsewhere.
\section {The Elastic Region}
\par
\begin{figure}[htb]
\begin{center}
\mbox{\epsfig{file=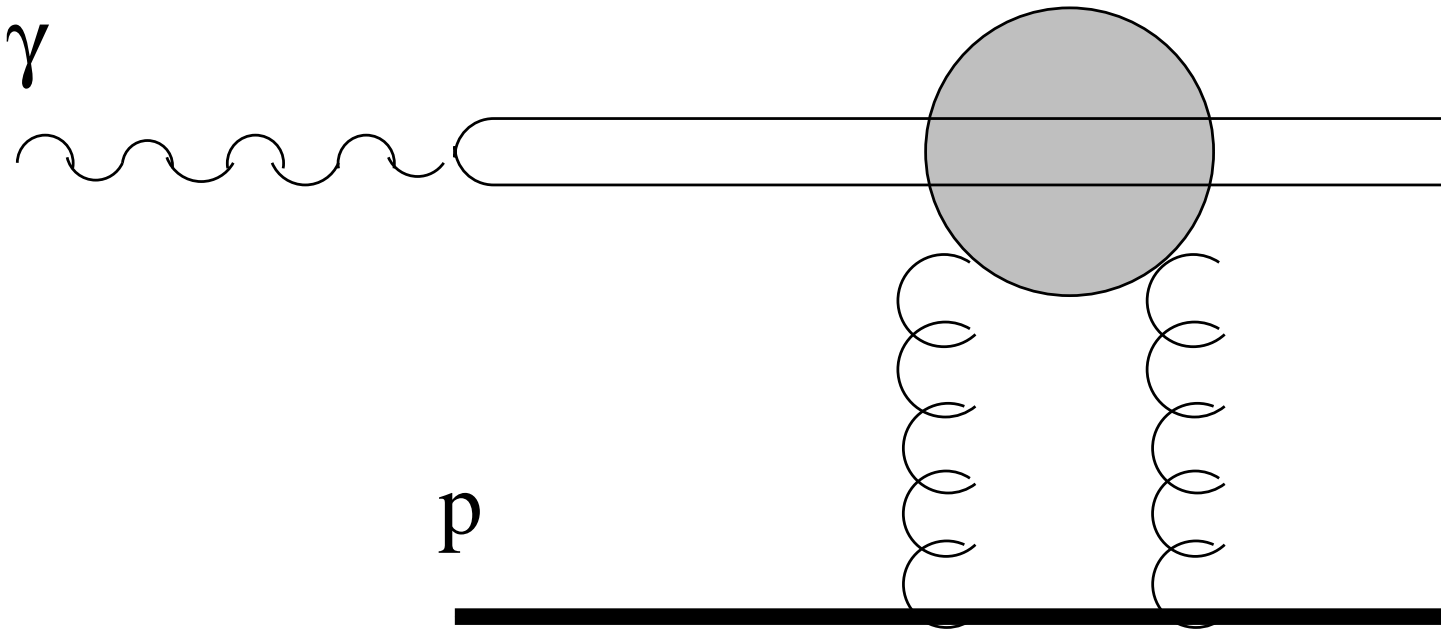,width=9truecm,angle=-90}}
\caption{ Diffractive \jpsi photoproduction.}
\label{Fig 5.}
\end{center}
\end{figure}
Having looked at \jpsi production in the inelastic limit, we now move
over to the diffractive region, i.e for values of the elasticity parameter
$z$ close to 1. The description of the scattering 
process via the standard photon gluon fusion
mechanism breaks down in this limit, as the elasticity parameter takes
values close to $1$, and the formulae of the previous section are no
longer applicable. In this kinematic regime of diffractive scattering,
BFKL dynamics takes over, and in particular, one has to sum over the
gluon ladder diagrams as shown in Fig(5).
In this elastic region, the
scattering amplitude occurs in a factorized form, \cite{rys},\cite{M/rys}  
which is essentially due to the fact
that in the proton rest frame, the formation time of the \jpsi is much
greater than the time of interaction of the photon with the proton. The
photoproduction amplitude with the incident photon almost on shell
is described by a two gluon exchange scattering process where
the transverse momentum transfer $t \approx 0$ for the diffractive
process. The amplitude for the process in [Fig. 5] was evaluated in the
leading log approximation in perturbative QCD in \cite{rys} and 
depends on the {\it square} of the gluon distribution function.
Subsequently, this formula has been improved upon in \cite{M/rys} where
the transverse momentum of the gluon ladder is included in the analysis,
thus incorporating BFKL \cite{bfkl} dynamics. Dependence on the square
of the gluon distribution function makes this cross section extremely
sensitive on the latter and a precise measurement of the cross section
will shed light on the behaviour of the gluon at very small
values of x. It may be mentioned that here we are using the DGLAP
evolved gluons in the double scaling limit as a mere parametrization of
the data, and hence it is not inappropriate to use them in a regime
where BFKL dynamics plays an important role.
In \cite{rys} the amplitude for the Feynman graphs of [Fig. 5] were
evaluated and found to be
\begin{equation}
{d \sigma (\gamma p \rightarrow \jpsi +p) \over dt}~=~[F]^{2}(t) {\Gamma
^{J}_{ee} \alpha_{s}^{2} m_{J}^{3} \pi^{3} \over 192 \alpha} \left[ {\bar
x} g( {\bar x}, {\bar Q}^{2}) {2 {\bar Q}^{2}~-~|Q_{T}|^{2} \over 
({ \bar Q}^{2})^{2}} \right]^{2} 
\end{equation}
where $\alpha$, $\alpha_{s}$ and $\Gamma^{J}_{ee}$ denote the
electromagnetic coupling, the strong coupling and the electronic width
of the \jpsi as before and the quantity $F$
denotes the two gluon form factor that is apriori unknown. To the first 
approximation it can be treated as the electromagnetic form factor, but
its nature is strictly determined experimentally. 
We shall however restrict ourselves to the domain of photoproduction  
of \jpsi in the diffractive regime, i.e where the photon is almost on
shell and the momentum transferred, $t$ is extremely small, $t \simeq 0$.
At such vanishingly small values of $t$, this form factor can be
approximated to unity \cite{rys}. 
Here, $Q_{T}$ denotes the transverse momentum of
the \jpsi particle, which is typically small, $\simeq 0.3GeV$. The
quantities $\bar x$ and $\bar Q^{2}$ is defined as
\begin{equation}
{\bar Q}^{2}~=~{ 1 \over 4}\left( |Q^{2}|+m_{J}^{2} \right);~~~~{\bar x}
~=~4{{\bar Q}^{2} \over S}  
\end{equation}
where $S$ is the photon-proton centre of mass energy.
As we wish to consider the photoproduction limit, we set the photon
virtuality to be extremely small, and in our calculations, we have
chosen the scale $\bar Q^{2} = 2.4 GeV^2$
In this expression, only
longitudinally polarised gluons have been taken into account as these
are the only ones that contribute in the extremely high c.m energy
limit. We note, first of all, that the dependence of the cross section on
the barred quantities ensure that perturbative QCD can be applied even
in the photoproduction limit, due to the heavy mass of the \jpsi.
Secondly, the expression for $x$ tells us that at sufficiently large
values of $S$, the extremely small limit of $x$ can be probed. 
\par
This formula (9), however has been derived in the leading ln $Q^{2}$
approximation, without the inclusion of the transverse momenta of the
exchanged gluons. 
The analysis of the above process including the gluon transverse momenta
has been systematically carried out in \cite{M/rys}. In their
calculation, the cross section is expressed in terms of the unintegrated
gluon distribution, $f_{BFKL}(x,k_{T}^{2})$, which satisfies the BFKL equation,
and is related to the conventional integrated gluon density by the
relation
\begin{equation}
xg(x,Q^{2})~=~\int^{Q^{2}} {dk_{T}^{2} \over k_{T}^{2}}f_{BFKL}(x,k_{T}^{2})
\end{equation}
Inclusion of the gluon $k_{T}^{2}$ modifies the scattering cross section
to 
\begin{equation}
{d \sigma (\gamma p \rightarrow \jpsi +p) \over dt}~=~{ |A|^{2} \over 16
\pi}
\end{equation}
with,
\begin{equation}
A~=~i 2 \pi^{2} m_{J} \alpha_{s} B \int {dk_{T}^{2} \over k_{T}^{2}}
\left({k_{T}^{2} \over ({\bar Q}^{2})({\bar Q}^{2}~+~\kts)} \right)
{dxg(x, \kts) \over d \kts }
\end{equation}
and $B^{2}~=~{\Gamma_{ee}^{J}m_{J} \over 12 \alpha}$
The amplitude $A$ 
can be written in terms of the conventional gluon distribution as
\begin{equation}
A~=~i2 \pi^{2} m_{J}  \alpha_{s} B \left[ {xg(x,Q_{L}^{2}) \over {\bar
Q}^{4}}~+~\int_{Q_{L}^{2}}^{\infty} {d \kts \over ({\bar Q}^{2})({\bar
Q}^{2}~+~\kts)}{dxg(x, \kts) \over d \kts }\right]
\end{equation}
The lower cutoff $Q_{L}$ has to be put on the integration due to the
undetermined form of the gluon distribution function at vanishingly
small values of \kts. This form of $A$ can be plugged in (12) in
order to obtain the total cross section.
\par
\begin{figure}[htb]
\begin{center}
\mbox{\epsfig{file=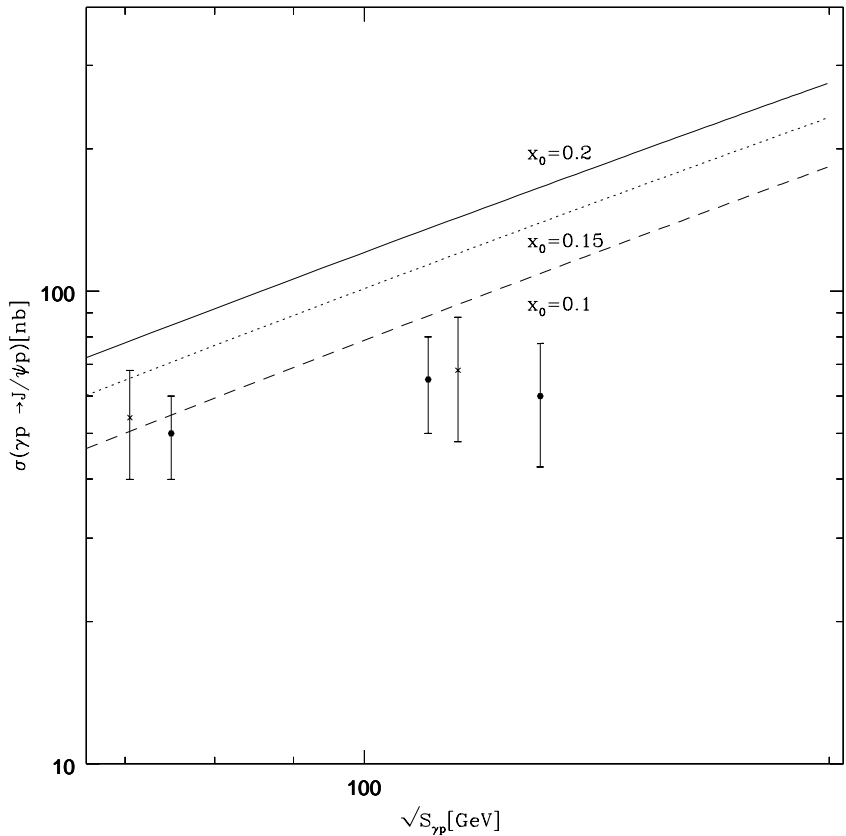,width=12truecm}}
\caption{Total cross section for diffractive \jpsi photoproduction. The
lower, middle and upper lines corresponds to $x_{0}=0.1, 0.15$ and $0.2$
respectively.}
\label{Fig 6.}
\end{center}
\end{figure}
We have obtained the values for the total cross section for diffractive
photoproduction of \jpsi mesons, using (12) and (14). 
From our analysis, it seems that using the soft 
pomeron normalisation in this region a la Ball and Forte \cite{bf}, the 
double scaling
analysis grossly overestimates presently available data, by a factor of 
about 5.
Hence, we conclude that in this region of extremely smally values of $x$, 
$x \leq 10^{-3}$, there is
some admixture of the hard pomeron boundary condition, and consequently to 
analyse
this region, we have chosen a normalisation corresponding to some 
intermediate pomeron, namely, $N=1.45$. 

As a lower cutoff for the
transverse momenta of the gluon distribution function, we have used the
value $Q_{L}^{2}=1.5GeV^{2}$. This value of $Q_{L}^2$ has been chosen so 
that it is just
above the starting scale for the gluon distribution (namely $Q_0^2=1.12GeV^2$)
in the double scaling analysis. We will come back to the issue of
variation of the cross section with this cutoff in a moment.
Note that in our calculation, we have
considered the forward scattering amplitude, namely $t=0$. There may be
corrections coming from consideration of a non zero value of the
momentum transfer, $|t_{min}|$; however, these corrections are expected
to be small in the region of $x$ that we are interested in, namely $
x<0.01$ \cite{M/rys}. One might also consider the corrections  
arising out of the relativistic effects of the \jpsi wave function, 
those coming from treating the quarks in the \jpsi bound state non
relativistically, i.e assuming $m_{J}=2m_{c}$, where $m_{c}$ is the mass
of the charm quark, and also those arising out of radiative corrections to the
lowest order process considered here.
Analysis of these correction factors have been
carried out in \cite{M/rys} and it has been shown that 
whereas the radiative correction is negligible for soft gluon emission,
as is the case here, the total correction factor arising from
relativistic effects in the \jpsi wave function and the motion of the
quarks inside the charmonium bound state is of the order of unity.
\par
The results of our analysis is shown in Fig [6]. 
From the available data points \cite{aid}, we find that 
the double scaling form of the gluon distribution function predicts a  
slightly steeper rise in the elastic cross section than favoured by the
experimental data. However, experimental values for larger centre of
mass energies than are currently probed may throw light on the scenario
for diffractive \jpsi photoproduction. 
We have also plotted the variation of the cross section with the choice
of the starting scale  
$x_{0}$ in the gluon distribution function in the double scaling
limit. It can be seen that change of $x_{0}$ causes a considerable
amount of fluctuation in the \jpsi production cross-section. 
We have checked that there is similar fluctuation in the cross section
as one varies the starting scale for the gluon distribution function,
namely $Q_{0}^2$, though the results are not explicitly indicated here.
We have also explored the sensitivity of the cross section to the lower
cutoff in the transverse momentum ($Q_L$) used to compute it.
It is found that the cross section is highly sensitive to fluctuations in
$Q_L^2$, varying from about $70 \% $ at $\sqrt S=100GeV$ 
to nearly $75 \%$ at $\sqrt
S=200GeV$. This is in contrast to the smaller variation in cross section noted
in \cite{rys} using the GRV and MRS gluons, as  $Q_L^2$ is varied.

\section{Conclusions}
In this paper, we have studied the validity of the double scaling ideas
to a specific semi-inclusive process like \jpsi production. 
Our results show that for the inelastic region, the data is
fairly well accounted for by the double scaling form of the gluon
distribution function, even in a region beyond that confirmed by
measurement of the proton structure function $F_2$. However, in the
diffractive scattering regime, double asymptotic scaling while predicting
a steeper increase in the cross section than from present data indicates the
admixture of hard pomeron boundary conditions (as opposed to the soft boundary
conditions \cite{bf} that we assumed while analysing the inelastic scattering 
data). As such, it is not very clear, at least  
from the present analysis as to which behaviour takes over at extremely
small values of $x$, and diffractive \jpsi photoproduction may 
provide some crucial
hints, as it is extremely sensitive to the gluon 
distribution function.

To our
knowledge, this is the first attempt to apply the idea of double scaling
to a semi-inclusive process. The advantage of using double scaling, as
pointed out in the introduction is that it does not need any given form
form for the starting gluon distribution $g(x_0, Q_0)$ as with other
parametrisations. We have studied whether the region
of validity of double scaling in a totally inclusive process like the
$F_{2}$ measurement changes when a semi-inclusive process like \jpsi
photoproduction is analysed. In the case of diffractive scattering, we
have also used the data to constrain the starting  value of $x$, i.e
$x_0$ and found significant change in the value of the total cross
section, a similar change being present if one varies the  
starting scale for the gluon distribution function, i.e $Q_0^2$. 

There have been many suggestions,
including a very elegant one by Mueller \cite{Mueller} for studying processes
that would clearly distinguish QCD evolution through DGLAP and some
other mechanism like BFKL. The basic thrust of all these suggestions has
been to study semi-inclusive processes and we have taken a first step in
this direction by studying the validity of double scaling in \jpsi
photoproduction where data with reasonably good statistics is already
available from HERA. It would be useful to make a systematic study of
all such semi-inclusive processes for which data exists, in different
resummation schemes (DGLAP, BFKL, CCFM, etc) in order to look for the
validity of these different methods. This would then give us a
qualitative understanding of the physics of the different processes and
consequently that of the various resummation schemes. These are some of
the issues for future study.

\vskip 1cm
\noindent{\bf \large Acknowledgements}\\[0.5cm]
T.S wishes to thank Mohan Narayan and 
K. Sridhar for lot of stimulating discussions.
R.B. would like to thank D. Indumathi for many helpful discussions and comments.
Both of us wish to thank the International Centre for Theoretical
Physics (ICTP), Trieste, Italy for its hospitality 
where the initial part of this work was
done.


\begin{thebibliography}{99}
\bibitem{bf}
R. D. Ball and S. Forte, Phys. Lett. {\bf B 335} (1994) 77; Phys.
Lett. {\bf B 336} (1994) 77.
\bibitem{bfkl}
E. A. Kuraev, L. N. Lipatov and V. S. Fadin, Zh. Eksp. Teor. Fiz {\bf
72} (1977) (Sov. Phys. JETP {\bf 45} (1977) 199); Ya. Ya. Balitzkij
and L. N. Lipatov, Yad. Fiz {\bf 28} (1978) 1597 (Sov. J. Nucl. Phys.
{\bf 28} (1978) 822); J. B. Bronzan and R. L. Sugar, Phys. Rev. {\bf
D17} (1978) 585; T. Jaroszewicz, Acta. Phys. Polom. {\bf B11} (1980)
965.
\bibitem{aid}
S. Aid et al, (H1 Collaboration) hep-ex/9603005, Nucl. Phys {\bf B 472} 
(1996) 3.
\bibitem{ap}
G. Altarelli and G. Parisi, Nucl. Phys. {\bf B 126} (1977) 278.  
\bibitem{aid1}
S. Aid et al, (H1 Collaboration) hep-ex/9603004, Nucl. Phys. {\bf B 470} (1996) 3.
\bibitem{B/J}
E.I. Berger and D.Jones, Phys. Rev. {\bf D 23}, (1981) 1521.
\bibitem{RMG}
R.M. Godbole, D.P. Roy, and K. Sridhar, Phys. Lett. {\bf B 373}, (1996)
328.
\bibitem{kram}
M. Kramer, hep-ph/9508409, Nucl. Phys {\bf B 459} (1996) 3.
\bibitem{mangano}
M. Mangano, CDF Collaboration, presented at 27th International
Conference on High Energy Physics, Glasgow, July(1994).
\bibitem{bratflem}
E. Braaten and S. Fleming, Phys. Rev. Lett {\bf 74} (1995) 3327.
\bibitem{bbl}
G.T. Bodwin, E. Braaten and G.P Lepage, Phys. Rev. {\bf D 51}, (1995)
1125.
\bibitem{cacc}
M. Cacciari and M. Kramer, hep-ph/9601276.
\bibitem{ng}
A. D. Martin, C. K. Ng and W. J. Stirling, Phys. Lett. {\bf B191} (1987)
200.
\bibitem{B/Y}
E. Braaten and T.C. Yuan, Phys. Rev. Lett. {\bf 71}, (1993) 1673.
\bibitem{B/Y1}
E. Braaten and T.C. Yuan, Phys. Rev. {\bf D 50}, (1994) 3176 ; E.
Braaten, K. Cheung and T.C. Yuan, Phys. Rev. {\bf D 48}, (1993) 4230;
Y.Q. Chen, Phys. Rev. {\bf D 48}, (1993) 5181; T.C. Yuan, Phys. Rev.
{\bf D 50}, (1994) 5664.
\bibitem{rys}
M.G Ryskin, Z. Phys {\bf C57}, (1993) 89.
\bibitem{M/rys}
M.G. Ryskin, R.G. Roberts, A.D. Martin and E.M. Levin, 
hep-ph/9511228.
\bibitem{Mueller}
A. H. Mueller, Nucl. Phys. {\bf B} (Proc. Suppl.), {\bf 18C},(1990) 125 ;
J. Phys. {G 17}, (1991) 1443.
\end{thebibliography}
\end{document}